\documentclass[epj,referee]{svjour}
\usepackage{epsfig}

\def\Q2tilde{\tilde{Q}^2}
\def\MSbar{$\overline{\mathrm{MS}}\ $}
\def\msbar{\overline{\tiny \mathrm{MS}}}

\def\ba{\begin{eqnarray}}
\def\ea{\end{eqnarray}}
\def\dd{{\mathrm d}}
\def\ii{{\mathrm i}}

\def\fun#1#2{\lower3.6pt\vbox{\baselineskip0pt\lineskip.9pt
  \ialign{$\mathsurround=0pt#1\hfil##\hfil$\crcr#2\crcr\sim\crcr}}}
\def\order#1{{\mathcal O}\left(#1\right)}

\newcommand{\nll}{\nonumber\\}
\newcommand{\sss}[1]{\scriptscriptstyle{#1}}
\def\GF {G_{\sss F}}
\def\gw {\Gamma_{\sss W}}
\def\gz {\Gamma_{\sss Z}}
\def\mw {M_{\sss W}}
\def\mz {M_{\sss Z}}
\def\mh {M_{\sss H}}

\newcommand{\GeV}{\unskip\,\mathrm{GeV}}
\newcommand{\MeV}{\unskip\,\mathrm{MeV}}

\begin{document}

\title{One-loop corrections to the Drell-Yan process in SANC} 
\subtitle{(I) The charged current case.}

\author{A. Arbuzov\inst{1} \email{arbuzov@thsun1.jinr.ru}
\and D. Bardin\inst{2} 
\and S. Bondarenko\inst{1} 
\and P. Christova\inst{2} 
\and L. Kalinovskaya\inst{2} 
\and G. Nanava\thanks{On leave from IHEP, TSU, Tbilisi, Georgia.}\inst{2}
\and R. Sadykov\inst{2}
}

\institute{Bogoliubov Laboratory of Theoretical Physics, 
JINR,\ Dubna, \ 141980 \ \  Russia
\and
Dzhelepov Laboratory of Nuclear Problems, 
JINR,\ Dubna, \ 141980 \ \  Russia}

\abstract{
Radiative corrections to the charged current Drell-Yan processes are revisited.
Complete one-loop electroweak corrections are calculated within 
the automatic SANC system. 
Electroweak scheme dependence and the choice of the factorization 
scale are discussed. Comparisons with earlier calculations are presented. 
}


\maketitle

\section{Introduction}

Precision studies of the Drell-Yan process are vitally important for
high energy hadronic colliders. This process provides information
about weak interactions and contributes to the background to many
of the searches for physics beyond the Standard Model.
One-loop QED and electroweak (EW) radiative corrections (RC) to the 
Drell-Yan process at high energy
hadronic collider were calculated by several groups in the past, see
papers~\cite{Mosolov:1981xk,Soroko:1990ug,Wackeroth:1996hz,Baur:1998kt,Dittmaier:2001ay,Baur:2004ig}
and references therein.
Here we present the results for the corrections to the charged current 
Drell-Yan process, obtained within the automatized system 
SANC~\cite{Andonov:2004hi,SANCwww} and some comparisons with  
earlier calculations. Starting from the construction of helicity amplitudes and
EW form factors, SANC performs calculation of the process cross section
and produces computer codes, which can be further used in the experimental
data analysis.

\section{Preliminaries and Notation}

Let us start with the partonic level, where we will consider interactions of
{\em free} quarks (partons). 
The differential Born-level cross section of the process
\ba \label{udlnu}
\bar{d}(p_1)\ +\ u(p_2)\ \to\ l^+(p_4)\ +\ \nu_l(p_3)
\ea
in the center-of-mass system of the initial quarks reads
\ba
&& \frac{\dd\hat\sigma_0}{\dd \hat\Omega} = \frac{1}{4}\frac{1}{N_c}|V_{ud}|^2
\frac{G_F^2\mw^2}{2\pi \hat{s}}\;
\frac{\hat{u}^2}{(\hat{s}-\mw^2)^2 + \gw^2(\hat{s})\mw^2},
\nonumber \\ && \qquad 
\hat{s} = (p_1+p_2)^2, \qquad 
\hat{u} = (p_1-p_3)^2,
\ea
where $N_C=3$ is the number of quark colors; $V_{ud}$ is the relevant element of 
the CKM matrix;
$G_F$ is the Fermi coupling constant; $\mw$ and $\gw$ are the mass and the width of
the $W$-boson, respectively.

\section{Radiative Corrections at the Partonic Level}

In order to get a more accurate description of the process we
should go beyond the Born approximation and take into account
different sources of radiative corrections. Here  we
will consider only EW contributions to the corrections, while
effects of higher order QCD contributions (and {\em mixed} effects)
are left beyond the scope of our study.

As usually, we subdivide the EW RC into the virtual (loop) ones,
the ones due to soft photon emission, and the ones due to hard 
photon emission. An auxiliary parameter $\bar\omega$ separates
the soft and hard photonic contributions. 

In the automatized system~\cite{SANCwww},
the virtual corrections are accessible via menu chain {\bf SANC $\to$ EW
$\to$ Processes $\to$ 4 legs $\to$ 4f $\to$ Charged Current $\to$
f1~f1'~$\to$~f~f'~(FF)}. The module, loaded at the end of this chain computes
on-line the scalar form factors of the partonic sub-process~(\ref{udlnu}). The parallel
module {\bf \dots f1~f1'~$\to$~f~f'~(HA)} provides the relevant helicity amplitudes.
For more details see Section~2.5 of the SANC description~\cite{Andonov:2004hi} 
and the book~\cite{Bardin:1999ak}.

The real photon emission process 
\ba
\bar{d}(p_1)\ +\ u(p_2)\ \to\ l^+(p_4)\ +\ \nu_l(p_3)\ +\ \gamma(p_5)
\ea
should be taken into account as well.
Integration over the phase space in this case can be performed either
(semi-)analytically or by means of a Monte Carlo integrator.

The first possibility is realized within the SANC environment.
Now we have there two branches. The first one contains the complete
chain of analytical integrals over the hard photon phase space. 
It provides at the partonic level the double-differential distribution 
$\dd^2\hat{\sigma}_{\mathrm{hard}}/(\dd c\; \dd \hat{s}')$ and
the single differential distribution 
$\dd\hat{\sigma}_{\mathrm{hard}}/\dd c$, where $c=\cos\angle(\vec{p}_2\vec{p}_4)$
and $\hat{s}'=(p_3+p_4)^2$.
The second branch provides the double-differential distribution 
$\dd^2\hat{\sigma}_{\mathrm{hard}}/(\dd c\; \dd M_x^2)$, where 
$M_x^2=2p_3p_5$ which is directly related to the charged lepton
energy in the center-of-mass system of the initial quarks: 
\ba
\hat{E}_\mu=p_4^0=\frac{\hat{s}+m_l^2-M_x^2}{2\sqrt{\hat{s}}}.
\ea 

We managed also to obtain analytically the
hard photon contribution as the single differential distribution
$\dd\hat{\sigma}_{\mathrm{hard}}/\dd {\hat s'}$. In this case 
we can use a system of reference with the $z$-axis along 
the real photon momentum $\vec p_5$. 
There the integration over three angular variables
is rather easy and we have a possibility even to keep all the light 
masses exactly. Below we give the expression without mass terms 
because of its simplicity:
\ba 
&& \frac{\dd\hat{\sigma}_{\mathrm{hard}}}{\dd {\hat s'}} = 
\hat{\sigma}_0\, \frac{ \alpha}{2 \pi}
  ~\frac{1}{\hat{s}^2}~ \frac{1}{{\hat s}-{\hat s}'} \Biggl\{ 
\left[{\hat s}^2+{\hat s}'^2 \right]\Biggl[ Q^2_l \nll 
&&\ \times \left(\ln\frac{\hat{s}'}{m_l^2} - 1 \right) 
+  \frac{\hat{s}'}{\hat{s}}
            ~  \frac{(\hat{s}-\mw^2)^2+\gw^2\mw^2}
                    {(\hat{s}'-\mw^2)^2 + \gw^2\mw^2} 
\nll &&\ 
\times  \biggl[
             Q^2_u \left(\ln\frac{\hat{s}}{m_u^2} - 1 \right) 
           + Q^2_d \left(\ln\frac{\hat{s}}{m_d^2} - 1 \right) \biggr]\Biggr] 
\nll &&\ 
- \frac{2}{3} \left({\hat s}^2+{\hat s}{\hat s}'+{\hat s}'^2\right) 
       \biggl[ Q^2_l 
\nll &&\ 
+  \frac{\hat{s}'}{\hat{s}} \,
   \frac{(\hat{s}-\mw^2)^2+\gw^2\mw^2}
                {(\hat{s}'-\mw^2)^2 + \gw^2\mw^2} \biggr] 
\nll &&\  
- \frac{1}{3} \left[{\hat s}'\left({\hat s}+{\hat s}'\right) \right]  
        Q_l   \Bigl( 4 Q_u + 5 Q_d \Bigr)
\nll &&\ 
\times  \frac{\left(\mw^2- \hat{s}\right)
\left(\mw^2- \hat{s}'\right)+\gw^2\mw^2}
              {(\hat{s}'-\mw^2)^2 + \gw^2\mw^2} 
  \Biggr\}, 
\ea
where $Q_l$, $Q_u$, and $Q_d$ are the charges of the charged lepton,
up-quark and down-quark, respectively. 

The differential distributions of the tree-level radiative process 
$\bar{d}+u \to \mu^+ +\nu_\mu +\gamma$ were compared with the corresponding
distributions obtained by means of the CompHEP package~\cite{Boos:2004kh}.
Cross section distributions in the cosine 
of the outgoing muon angle and in the
muon energy are considered. 20 bins are constructed for each of the 
distributions. Bins in the muon energy are
\ba
(n_{\mathrm{bin}}-1)\ \times\ 5~\mathrm{GeV} < E_\mu 
< n_{\mathrm{bin}}\ \times\ 5~\mathrm{GeV}.
\ea
The cut on the muon energy ($E_\mu < 95$~GeV) is imposed in both distributions 
to avoid the region with soft photons, where CompHEP is not supposed to work well.
The angular bins are
\ba
-1 + \frac{n_{\mathrm{bin}}-1}{10} < c < -1 +  \frac{n_{\mathrm{bin}}}{10}\, .
\ea
The $\alpha(M_Z)$ electroweak scheme (realized according to the CompHEP conventions) 
was used. An agreement was found as can be seen from Table~\ref{Table0}. 

For the two choices of variables we have simple
analytical expressions for the corresponding soft photon contributions.
The infrared singularities in them are regularized by the auxiliary photon 
mass. The energy of a soft photon is limited from above by a cut 
in the integral either over $\hat{s}'$ or over $M_x^2$. 

In order to have the possibility to impose experimental cuts and
event selection procedures of any kind, we can use a Monte Carlo
integration routine based on the Vegas algorithm~\cite{Lepage:1977sw}.
In this case we perform a 4(6)-fold numerical integration to get the
hard photon contribution to the partonic (hadronic) cross section. 
To get the total EW correction we add also the contributions 
of the soft photon 
emission and the ones of the virtual loops. The  cancellation of the 
dependence on the auxiliary parameter $\bar\omega$ 
in the sum is observed numerically.

Using the splitting of the $W$-boson propagators in the case of real
photon emission off the virtual $W$, we separate the contributions of the 
initial state radiation, the final state one, and their interference 
in a gauge invariant way~\cite{Berends:1984qa}. 
The splitting is introduced by the following formula:
\ba
&& \frac{1}{\hat{s}-(\mw-\ii\gw)^2}\cdot
\frac{1}{\hat{s}'-(\mw-\ii\gw)^2} 
\nonumber \\ && \quad
= \frac{1}{(\hat{s}-\hat{s}')}
\biggl( \frac{1}{\hat{s}'-(\mw-\ii\gw)^2}
\nonumber \\ && \qquad
- \frac{1}{\hat{s}-(\mw-\ii\gw)^2}\biggr).
\ea
In the center-of-mass system $(\hat{s}-\hat{s}')=2p_5^0\sqrt{\hat{s}}$.
The fixed $W$-width scheme is used here and in what follows.

In the course of calculations of the $\order{\alpha}$ corrections we 
met the so-called {\em on-shell} singularities, which appear in the
form of $\ln(\hat{s}-\mw^2+i\epsilon)$. 
As was shown in detail in Ref.~\cite{Wackeroth:1996hz}, they can be 
regularized by the $W$-width:
\ba
\ln(\hat{s}'-\mw^2+i\epsilon) \to \ln(\hat{s}'-\mw^2+i\mw\gw).
\ea

In the analytical formulae for radiative corrections one can find
logarithms with quark and lepton mass singularities:
\ba
\ln\frac{\hat{s}}{m_l^2}\, , \qquad
\ln\frac{\hat{s}}{m_u^2}\, , \qquad
\ln\frac{\hat{s}}{m_d^2}\, .
\ea
In the experimental set-up with calorimetric registration
of the final state charged particles (typical for electrons), 
the lepton mass singularity cancels out in the result for 
the correction to an observable cross section in accordance with 
the Kinoshita--Lee--Nauenberg theorem~\cite{Kinoshita:1962ur,Lee:1964is}.
But if the experiment is measuring the energy of the charged lepton
without summing it with the energies of accompanying collinear photons 
(typical for muons), the logarithms with the lepton mass singularity 
remain in the result and give a considerable numerical contribution. 
Re-summation of these logs in higher orders was discussed 
in Refs.~\cite{Placzek:2003zg,CarloniCalame:2003ux,CarloniCalame:2003ck}.

\subsection{Treatment of Quark Mass Singularities
\label{qms}}

One-loop radiative corrections contain terms proportional to the
logarithms of the quark masses, $\ln(\hat{s}/m_{u,d}^2)$. 
They come from the initial state radiation contributions including hard, 
soft and virtual photon emission. Such initial state mass singularities 
are well known, for instance, in the process of $e^+e^-$ annihilation.
But in the case of hadron collisions these logs have been already 
{\em effectively} taken into account in the parton density functions (PDF's).
In fact, in the procedure of PDF's extraction from the experimental data,
QED radiative corrections to the quark line have not been systematically
subtracted. Therefore the present PDF's effectively include
not only the QCD evolution but also the QED one. 
Moreover, it is known that the leading log behaviors of 
the QED and QCD DGLAP evolution of quark density functions are
similar (proportional to each other). So one gets evolution
of PDF's with an effective coupling constant 
\ba
\alpha^{\mathrm{eff}}_{s} \approx \alpha_{s} + \frac{Q_i^2}{C_F}\alpha,
\ea
where $\alpha_s$ is the strong coupling constant, $\alpha$ is the fine structure 
constant, $Q_i$ is the quark charge, and $C_F$ is the QCD color 
factor. The nontrivial difference between the QED evolution and the QCD one 
starts to appear in higher orders,
and the corresponding numerical effect is small compared to the remaining QCD
uncertainties in PDF's~\cite{Kripfganz:bd,Spiesberger:1994dm,Roth:2004ti,Martin:2004dh}.
The best approach to the whole problem would be to re--analyze  all
the experimental DIS data taking into account QED corrections to the quark line
at least at the next--to--leading order. But for the present moment we can limit
ourselves with an application of a certain subtraction scheme
to the QED part of the radiative corrections for the process under consideration. 
We will use here the \MSbar scheme~\cite{Bardeen:1978yd}, the DIS scheme can
be used as well.  This allows to avoid the double counting of the 
initial quark mass singularities contained in our
result for the corrections to the free quark cross section
and  the ones contained in the corresponding PDF. 
The latter should be also taken 
in the same scheme with the same factorization scale.

In fact, using the initial condition for the non--singlet NLO QED quark 
structure function,
which coincides with the QCD one with the trivial substitution 
$C_F\alpha_s\to Q_i^2\alpha$, see Ref.~\cite{Berends:1987ab}, 
one gets the following expression for the terms to be subtracted from the full 
calculation with massive quarks:
\ba 
\delta^{\msbar}  &=& \sum\limits_{i=1,2}^{}
Q_i^2 \frac{\alpha}{2\pi} \int\limits_0^1 \dd \xi_i\;  \biggl[
\frac{1+\xi_i^2}{1-\xi_i} \biggl( \ln\frac{M^2}{m_i^2} 
\nonumber \\
&-& 2\ln(1-\xi_i) - 1 \biggr) \biggr]_+ \hat\sigma_0(\xi_i),
\ea
where $Q_i$ and $m_i$ denote the charge and the mass of the given quark;
$M$ is the factorization scale; $\hat\sigma_0(\xi_i)$ is the  
cross section at the partonic level with the reduced value of the quark
momentum: $p_i\to \xi_i p_i$. The subtracted partonic cross 
section with $\order{\alpha}$
corrections is given by
\ba \label{msbarsi}
\hat{\sigma}_1^{\msbar} = \hat{\sigma}_1 - \delta^{\msbar}.
\ea
Then it can be convoluted with PDF's as shown below in Eq.~(\ref{sigpp}).

But there is an alternative way to perform the subtraction. Really, to
avoid the double counting of the quark mass singularities, we can leave them
in the corrected cross section, but remove from the PDF's:
\ba \label{msbarq}
&& \bar{q}(x,M^2) = q(x,M^2) -
\int_x^1 \frac{\dd z}{z} \, q\biggl(\frac{x}{z},M^2\biggr) \,
\frac{\alpha}{2\pi} \, Q_q^2 
\nonumber \\ && \quad \times
\biggl[ \frac{1+z^2}{1-z}
\biggl\{\ln\biggl(\frac{M^2}{m_q^2}\biggr)-2\ln(1-z)-1\biggr\} \biggr]_+
\nonumber \\ && \quad \equiv q(x,M^2) - \Delta q,
\ea
where $q(x,M^2)$ can be taken directly from the existing PDF's in the \MSbar
scheme (see Ref.~\cite{Wackeroth:1996hz} for the corresponding formula in the
DIS scheme). It can be shown analytically (see i.e. Ref.~\cite{Wackeroth:1996hz}), 
that this procedure is equivalent to the subtraction from the cross section,
and that it really removes (hides) the dependence on the quark masses. 
The advantage of the last approach is that it can be used regardless
of the way to represent the partonic cross section: it can be kept even
in the completely differential form.

The {\em natural} choices of the factorization scale are $M^2=\mw^2$ (when the
returning to the $W$-resonance is allowed by kinematic cuts) and
$M^2=\hat{s} = x_1 x_2 s$. Variations with respect to the choice 
should be studied.

In order to avoid the appearance of spurious higher order terms for the case
of subtraction from PDF's, we suggest to apply a procedure of {\em linearization}.
Schematically it can be represented as follows:
\ba \label{linea}
&& \bar{q}_1(x_1,M^2)\times \bar{q}_2(x_2,M^2)\times \hat{\sigma}_1 =
[q_1(x_1,M^2)-\Delta q_1]
\nonumber \\ && \quad 
\times [q_2(x_2,M^2)-\Delta q_2]\times
(\hat{\sigma}_{\mathrm{Born}} + \hat{\sigma}_{\alpha}) 
\nonumber \\ && \quad
\to q_1(x_1,M^2)\times q_2(x_2,M^2) \times \hat{\sigma}_{\mathrm{Born}} 
\nonumber \\ && \quad
+ q_1(x_1,M^2)\times q_2(x_2,M^2)\times \hat{\sigma}_{\alpha}
\\ \nonumber && \quad 
- [q_1(x_1,M^2)\times\Delta q_2 + q_2(x_2,M^2) \times\Delta q_1]
\times \hat{\sigma}_{\mathrm{Born}},
\ea
where $\hat{\sigma}_{\mathrm{Born}}$ and $\hat{\sigma}_{\alpha}$ denote the
Born-level partonic cross section and the $\order{\alpha}$ RC 
contribution to it, respectively. Without the linearization procedure,
terms with quark mass singularities would remain 
in the $\order{\alpha^2}$ contribution to the cross section.

\section{Radiative Corrections to Hadronic processes}

The double-differential cross section of the Drell-Yan process can
be obtained from the convolution of the partonic cross section with 
the quark density functions:
\ba \label{sigpp}
&& \frac{\dd\sigma_{\mathrm{RC}}^{pp\to\mu^+\nu X}(s)}{\dd c\;\dd E_\mu} 
= \sum\limits_{q_1q_2}\int\limits_{0}^{1} \int\limits_{0}^{1} 
\dd x_1\; \dd x_2\; \bar{q}_1(x_1,M^2) 
\nonumber \\ && \quad \times
\bar{q}_2(x_2,M^2)
\frac{\dd^2\hat{\sigma}^{q_1q_2\to\mu^+\nu}(\hat{s})}
{\dd\hat{c}\;\dd\hat{E}_\mu}\;{\mathcal J}\;\Theta(c,E_\mu),
\ea
where the parton densities with {\em bars} mean the ones modified by
the subtraction of the quark mass singularities;
the step function $\Theta(c,E_\mu)$ defines the phase space domain 
corresponding to the given event selection procedure.
The partonic cross section is taken in the center-of-mass reference 
frame of the initial quarks, where the cosine of the muon scattering
angle, $\hat{c}$, and the muon energy, $\hat{E}_\mu$, are defined.
The transformation into the observable variables $c$ and $E_\mu$ 
involves the Jacobian:
\ba
&& {\mathcal J} = \frac{\partial \hat{c}}{\partial c} \cdot
\frac{\partial \hat{E}_\mu}{\partial E_\mu} = \frac{4x_1x_2}{a^2}\cdot
\sqrt{\frac{a^2(1+c)}{x_1[a+x_2(1+c)]}}\,\, ,
\nonumber \\
&& a = x_1 + x_2 - c(x_1 - x_2), \quad \hat{c} = 1 - (1-c)\frac{2x_1}{a},
\nonumber \\ && 
\hat{s} = sx_1x_2,\qquad \hat{E}_\mu = \frac{\sqrt{\hat{s}}}{2}\, ,
\nonumber \\ && 
\hat{E}_\mu = E_\mu\sqrt{\frac{1-c^2}{1-\hat{c}^2}}\,\,  .
\ea
An analogous formula can be written for any other choice of a differential
distribution as well as for the total cross section.

\section{Numerical Results and Conclusions
\label{NumRes}}

For numerical evaluations we take the same set of input parameters as 
the one given by Eq.~(4.1) of Ref.~\cite{Dittmaier:2001ay}.
In Table~\ref{Table1} we present the results for the total
cross section\footnote{factor $|V_{ud}|^2$ has been dropped in the sake of comparison
with Ref.~\cite{Dittmaier:2001ay}.} 
of the process $u+\bar{d}\to \nu_l+l^+(+\gamma)$.
For the Born-level cross section we completely (in all listed digits) agree with the
numbers given in Ref.~\cite{Dittmaier:2001ay}. The third line shows radiative corrections
in percent before the subtraction of quark mass singularities. These numbers were
received directly from the SANC system for the $G_F$ EW scheme. Starting from the fourth
line we use the treatment of the EW scheme\footnote{In $G_F'$ scheme we assigned 
the following one-loop value of the coupling constant standing at the photon vertices: 
$\alpha_{QED}\approx 1/132.544$.}, which has been adopted
Ref.~\cite{Dittmaier:2001ay}. The results for the radiative corrections
with \MSbar subtraction (with factorizations scale being equal to $M_W$) 
are also in a fair agreement. The small deviations there can be due to details in the
treatment of EW scheme with respect induced higher order effects. 
Huge positive corrections in the case without subtraction of quark mass singularities
above the $W$-peak are due to the initial state radiation which provides
the radiative return to the $W$-resonance.

The effect of EW scheme dependence is illustrated by Table~\ref{Table2}.
Results for the total partonic cross section at the Born and $\order{\alpha}$
levels are given for two EW schemes. At the Born level the $7.3\%$ difference
appears just due to the difference in the definition of EW constants in the
$G_F$ and in the $\alpha(0)$ schemes. As it should be the difference
between the corrected cross sections is less than the one at the Born level. 
But still it is large and comparable with the ordered precision of the calculation. 
Certainly, usage of the $\alpha(0)$ is not well motivated for the given energy range.
And the difference $\delta_1$ gives only an upper estimate of 
the  uncertainty due to the EW scheme dependence. In any case we are going to
perform further studies of this effect.
 
Table~\ref{Table3} represents the dependence of the hadronic Drell--Yan cross section
on the values of the quark masses with and without the subtraction procedure.
The conditions are as follows: the center-of-mass energy is 200~GeV, all events with the
invariant mass of the neutrino and charged lepton pair above $\sqrt{40}$~GeV are
accepted. $\sigma_0$ denotes the Born-level cross section obtained using the CTEQ4L
set of PDF's~\cite{Lai:1996mg}. $\sigma_1$, $\sigma_1^{\msbar(\sigma)}$, 
$\sigma_1^{\msbar(q)}$, and $\sigma_1^{\msbar(q)}$(lin.) 
stand for the cross sections with one-loop EW RC included. 
The double counting of the 
quark mass singularities in $\sigma_1$ is not removed. 
The \MSbar procedure~(\ref{msbarsi}) is applied to the partonic cross 
section in the computation of $\sigma_1^{\msbar(\sigma)}$. 
Values of $\sigma_1^{\msbar(q)}$ and 
$\sigma_1^{\msbar(q)}$(lin.) are computed by convolution of the quark (parton) 
density function modified according to Eq.~(\ref{msbarq}) with 
the full (including quark mass singularities) partonic cross section. 
The linearization procedure~(\ref{linea}) was adopted 
for $\sigma_1^{\msbar(q)}$(lin.) in addition. One can see that the numerical 
effect of linearization for the given set--up is small (but visible).
The two approaches to remove the double counting give very close results 
as it should be.

For an internal test of our calculations, a comparison of the results 
produced by our Monte Carlo (MC) and semi--analytical (SA) codes for the 
description of hard photon contributions was performed. The results are
presented in Table~\ref{Table4a}, where the
corresponding contributions to the proton--proton cross section at 14~TeV
center-of-mass energy are given. 
The conditions and the input parameters were the taken as the ones used 
in Ref.~\cite{tuned}:
\ba \nonumber
\begin{array}[b]{lcllcllcl}
G_F & = & 1.16637 \times 10^{-5} \GeV^{-2}, & && \\
\alpha(0) &=& 1/137.03599911, & 
\alpha_s &=& 0.1187, \\
\mw & = & 80.425\GeV, &
\gw & = & 2.124\GeV, \\
\mz & = & 91.1867\GeV,& 
\gz & = & 2.4952\GeV, \\
\mh & = & 150\GeV, &
m_t & = & 174.17\;\GeV, \\
m_u & = & m_d = 66\;\MeV, &
m_c & = & 1.55\;\GeV, \\
m_s & = & 150\;\MeV, &
m_b & = & 4.5\;\GeV, \\
|V_{ud}| & = & |V_{cs}| = 0.975, &
|V_{us}| & = & |V_{cd}| = 0.222. 
\end{array}
\ea
The MRST204QED set~\cite{Martin:2004dh} of PDF's and the $\GF$ EW scheme 
were used. Six values for the cut on the muon transverse momentum, $P_T$,
are considered. The cut on the muon rapidity is $|\eta_l|< 1.2$.
The cut on the missing momentum was not imposed, since it can't be 
realized in the semi-analytical branch.
We also show there the values of the total 
one-loop EW correction, $\delta\sigma^{MC,SA}_{tot}$.
The Table~\ref{Table4a} shows results for two values of the soft--hard
photon separator, $\bar\omega$, and justifies the independence of the total correction
on it within the accuracy achieved. The separator is defined in 
the center-of-mass reference frame of the colliding quarks (partons). 
We stress, that having a semi-analytical
branch of calculations served us as a benchmark and helped a lot to adjust 
the Monte Carlo code.

In this way with help of the automatized SANC system we calculated the
complete one-loop radiative corrections to the charged current Drell-Yan
cross section. Our results at the partonic level are in a good agreement
with the ones published earlier in Ref.~\cite{Dittmaier:2001ay}. 
The corresponding computer codes in analytical ({\tt FORM}) and numerical
({\tt FORTRAN}) formats are available from SANC~\cite{SANCwww}.
They can be used as a part of a more general computer program (like a
Monte Carlo event generator) to describe the Drell-Yan process in realistic
conditions. Further comparison at the hadronic level with analogous calculations
of other groups is in progress~\cite{tuned}.

\begin{acknowledgement}
We are grateful to C.~Carloni Calame,
S.~Dittmaier, S.~Jadach, M.~Kr\"amer, G.~Montagna,  O.~Nicrosini
W.~Placzek, A.~Vicini, Z.~Was for discussions.
Three of us (D.B., L.K. and G.N.) are indebted to the directorate 
of IFJ, (Cracow, Poland) for a hospitality extended to them in 
April--May 2005, when a part of this study was completed.
This work was supported by the INTAS grant 03-51-4007.
One of us (A.A.) thanks for support the grant of the Prezident RF
(Scinetific Schols 2027.2003.2) and the RFBR grant 04-02-17192.
\end{acknowledgement}


\onecolumn

\section{Tables}

\begin{table}[ht]
\centerline{
\begin{tabular}{|r|l|l||l|l|}
\hline
     &\multicolumn{2}{c||}{bins in $E_\mu$}
     &\multicolumn{2}{c| }{bins in $c$} \\ \hline
bin  &   SANC    &  CompHEP  &   SANC    &  CompHEP  \\ \hline
  1  & 0.0006(1) & 0.0006(1) & 0.5867(1) & 0.5869(2) \\
  2  & 0.0010(1) & 0.0010(1) & 0.3538(1) & 0.3537(1) \\
  3  & 0.0023(1) & 0.0023(1) & 0.1857(1) & 0.1858(1) \\
  4  & 0.0452(1) & 0.0451(1) & 0.1111(1) & 0.1111(1) \\
  5  & 0.0569(1) & 0.0569(1) & 0.0740(1) & 0.0741(1) \\
  6  & 0.0549(1) & 0.0549(1) & 0.0541(1) & 0.0541(1) \\
  7  & 0.0546(1) & 0.0546(1) & 0.0427(1) & 0.0428(1) \\
  8  & 0.0563(1) & 0.0563(1) & 0.0360(1) & 0.0361(1) \\
  9  & 0.0603(1) & 0.0603(1) & 0.0321(1) & 0.0321(1) \\
 10  & 0.0667(1) & 0.0666(1) & 0.0298(1) & 0.0297(1) \\
 11  & 0.0755(1) & 0.0755(1) & 0.0287(1) & 0.0287(1) \\
 12  & 0.0868(1) & 0.0867(1) & 0.0284(1) & 0.0285(1) \\
 13  & 0.1008(1) & 0.1008(1) & 0.0286(1) & 0.0287(1) \\
 14  & 0.1175(1) & 0.1175(1) & 0.0292(1) & 0.0292(1) \\
 15  & 0.1372(1) & 0.1372(1) & 0.0299(1) & 0.0299(1) \\
 16  & 0.1605(1) & 0.1605(1) & 0.0302(1) & 0.0301(1) \\
 17  & 0.1881(1) & 0.1881(1) & 0.0294(1) & 0.0293(1) \\
 18  & 0.2227(1) & 0.2227(1) & 0.0263(1) & 0.0263(1) \\
 19  & 0.2739(1) & 0.2740(2) & 0.0187(1) & 0.0187(1) \\
 20  & 0.0       & 0.0       & 0.0059(1) & 0.0059(1) \\
\hline
\end{tabular}
}
\vspace*{3mm}
\caption{Bin by bin comparison of differential distributions
for the process $\bar{d}+u \to \mu^++\nu_\mu+\gamma$.}
\label{Table0}
\end{table}

\begin{table}[ht]
\centerline{
\begin{tabular}{|c||c|c|c|c|c|c|c|}
\hline
$\sqrt{\hat s}/\mathrm{GeV}$ & 40 & 80 & 120 & 200 & 500 & 1000 & 2000 
\\  \hline
$\hat{\sigma}_0/\mathrm{pb}$ & 2.646 & 7991.4 & 8.906 & 1.388 & 0.165 & 
0.0396 & 0.00979
\\  \hline
$\delta/\%$,~full, $\GF$ &
$-1.70$&$-7.62$& 89.9  & 125.3  & 155.9  & 166.9 & 173.0
\\  \hline
$\delta/\%$,~full, $\GF'$ &
$-1.76$&$-7.87$& 92.9  & 129.6  &161.2   & 172.5 & 178.9
\\  \hline
$\delta/\%$,~\MSbar$\!\!(s)$, $\GF'$ &
  0.56 &  2.48 &$-17.2$&$-16.0$ &$-17.9$ &$-23.2$&$-31.5$
\\  \hline
$\delta/\%$,~\MSbar$\!\!(\mw$), $\GF'$ &
  0.73 &  2.48 &$-12.9$& $-3.2$ & 11.6   & 18.8  & 22.8  
\\  \hline
$\delta/\%$,~\cite{Dittmaier:2001ay}&
  0.7  &  2.42 &$-12.9$& $-3.3$ & 12     & 19    & 23
\\  \hline
\end{tabular}
} 
\vspace*{3mm}
\caption{The total lowest-order partonic cross section $\hat\sigma_0$ in the $G_F$ 
EW scheme and the corresponding relative one--loop correction $\delta$.}
\label{Table1}
\end{table}

\begin{table}[ht]
\centerline{
\begin{tabular}{|c||c|c|c|c|c|c|c|}
\hline
$\sqrt{\hat s}/\mathrm{GeV}$ & 40 & 80 & 120 & 200 & 500 & 1000 & 2000 
\\  \hline
$\hat\sigma_0/\mathrm{pb}$, $[G_F]$ 
& 2.646 & 7991.4 & 8.906 & 1.388 & 0.165 & 0.0396 & 0.00979
\\  \hline
$\hat\sigma_0/\mathrm{pb}$, $[\alpha(0)]$ 
& 2.454 & 7410.2 & 8.258 & 1.287 & 0.153 & 0.0368 & 0.00908
\\  \hline
$\delta_0/\%$~(diff)      &
$-7.3$&$-7.3$& $-7.3$&$-7.3$  &$-7.3$  & $-7.3$& $-7.3$
\\  \hline
$\hat\sigma_1/\mathrm{pb}$,~\MSbar($\mw$), $[G_F]$ 
& 2.665 & 8183.2 & 7.796 & 1.345 & 0.183 & 0.0467 & 0.01195
\\  \hline
$\hat\sigma_1/\mathrm{pb}$,~\MSbar($\mw$), $[\alpha(0)]$ 
& 2.617 & 8029.5 & 7.721 & 1.324 & 0.179 & 0.0455 & 0.01162
\\  \hline
$\delta_1/\%$~(diff)      &
$-1.8$&$-2.0$& $-0.5$&$-1.5$  &$-2.6$  & $-3.1$& $-3.3$
\\  \hline
\end{tabular}
}
\vspace*{3mm}
\caption{The total partonic cross section in the $G_F$ 
and $\alpha(0)$ EW schemes.}
\label{Table2}
\end{table}

\begin{table}[ht]
\centerline{
\begin{tabular}{|c|l|l|l|l|l|}
\hline
                   & $\sigma_0$ [pb] & $\sigma_1$ & $\sigma_1^{\msbar(\sigma)}$
& $\sigma_1^{\msbar(q)}$  & $\sigma_1^{\msbar(q)}$(lin.) 
\\  \hline
$m_u=m_d=4.85$~MeV & 2.5577(1) & 2.4795(3) & 2.5724(3) & 2.5704(3) & 2.5729(3)
\\  \hline
$m_u=m_d=48.5$~MeV & 2.5577(1) & 2.4992(3) & 2.5724(3) & 2.5713(3) & 2.5727(3)
\\  \hline
$m_u=m_d=485.$~MeV & 2.5577(1) & 2.5190(3) & 2.5724(3) & 2.5719(3) & 2.5726(3)
\\  \hline
\end{tabular}
}
\vspace*{3mm}
\caption{The tree level and corrected hadronic Drell--Yan cross section
different values of the light quark masses.}
\label{Table3}
\end{table}


\begin{table}[ht]
\centerline{
\begin{tabular}{|c|l|l|l|l|l|l|}
\hline
$p_T$ [GeV]     & $>25$     & $>50$    & $>100$    & $>200$    & $>500$      & $>1000$
\\  \hline
$\sigma_{Born}$ [pb] & 2112.22(2)&13.1507(2)&0.94506(1) &0.115106(1)&0.00548132(6)&0.000262108(3)
\\  \hline
     &\multicolumn{6}{c|}{$\bar\omega=0.01$~GeV}
\\  \hline
$\delta\sigma_{hard}^{MC}$ [\%] & 27.52(2) & 34.02(2) & 43.88(2) & 51.22(2) & 59.67(2) & 65.22(2) 
\\
$\delta\sigma_{hard}^{SA}$ [\%] & 27.54(2) & 34.02(2) & 43.87(2) & 51.21(2) & 59.68(2) & 65.22(2) 
\\  \hline
$\delta\sigma^{MC}_{tot}$ [\%] & $-$2.69(2) & $-$3.83(2) & $-$6.67(2) & $-$11.77(2) & $-$22.28(2) & $-$33.36(2) 
\\
$\delta\sigma^{SA}_{tot}$ [\%] & $-$2.68(2) & $-$3.83(2) & $-$6.68(2) & $-$11.78(2) & $-$22.27(2) & $-$33.36(2) 
\\  \hline
     &\multicolumn{6}{c|}{$\bar\omega=0.001$~GeV}
\\  \hline
$\delta\sigma_{hard}^{MC}$ [\%] & 36.85(2) & 44.35(2) & 55.70(2) & 64.25(2) & 74.14(2) & 80.72(2)   
\\
$\delta\sigma_{hard}^{SA}$ [\%] & 36.88(2) & 44.37(2) & 55.70(2) & 64.26(2) & 74.15(2) & 80.72(2)   
\\  \hline
$\delta\sigma^{MC}_{tot}$ [\%] & $-$2.70(2) & $-$3.85(2) & $-$6.69(2) & $-$11.76(2) & $-$22.29(2) & $-$33.35(2) 
\\
$\delta\sigma^{SA}_{tot}$ [\%] & $-$2.67(2) & $-$3.83(2) & $-$6.69(2) & $-$11.76(2) & $-$22.28(2) & $-$33.35(2)   
\\  \hline
\end{tabular}
}
\vspace*{3mm}
\caption{The hadronic Drell--Yan cross section in the Born approximation 
and the hard photon contributions to it 
for different values of the cut on the muon transverse momentum.}
\label{Table4a}
\end{table}

\end{document}